\RequirePackage{lineno}  
\documentclass[twoside,slac_one]{revtex4}
\usepackage{graphicx}
\usepackage{fancyhdr}
\usepackage{amsmath} % American Mathematics Society standards
\usepackage{bm}% bold math
\usepackage{amsxtra}
\usepackage{amssymb}
\usepackage{amsthm}
\usepackage{latexsym}
\usepackage{lscape}
\usepackage{natbib}

\def\simge{\mathrel{%
   \rlap{\raise 0.511ex \hbox{$>$}}{\lower 0.511ex \hbox{$\sim$}}}}
\def\simle{\mathrel{
   \rlap{\raise 0.511ex \hbox{$<$}}{\lower 0.511ex \hbox{$\sim$}}}}
\newcommand{\mH}{m_H}
\newcommand {\GeV} {{\rm GeV}}
\newcommand{\HWWstarlvlv}{H\rightarrow WW^{(\ast)}\rightarrow \ell\nu\ell\nu}
\newcommand{\HWWlvlv}{H\rightarrow WW\rightarrow \ell\nu\ell\nu}
\newcommand{\HWWlvqq}{H\rightarrow WW\rightarrow \ell\nu jj}
\newcommand{\MET}{\mbox{$E\kern-0.60em\raise0.10ex\hbox{/}_{T}$}}

\begin{document}

\bibpunct{[}{]}{;}{n}{,}{,}

\title{Search for the Standard Model Higgs Boson in the Lepton + Missing
  Transverse Energy + Jets Final State in ATLAS}


\author{M.S. Neubauer on behalf of the ATLAS Collaboration}
\affiliation{Department of Physics, University of Illinois at
  Urbana-Champaign, Urbana, IL, USA}

\begin{abstract}
A search for the Standard Model Higgs boson has been performed in the
$H\rightarrow WW\rightarrow\ell\nu jj$ channel in 1.04 ${\rm fb}^{-1}$
of $pp$ collisions at $\sqrt{s} = 7~{\rm TeV}$ collected with the
ATLAS detector at the Large Hadron Collider. No significant excess of
events is observed over the expected background and limits on the
Higgs boson production cross section are derived for a Higgs boson
mass in the range $240~{\rm GeV}<\mH<600$~GeV. The best sensitivity is
reached for $\mH = 400~\GeV$, where the 95\% confidence level upper
bound on the cross-section for Higgs boson production times the
branching ratio for $H\rightarrow WW$ is 3.1 pb, or 2.7 times the
Standard Model prediction.
\end{abstract}

\maketitle

\thispagestyle{fancy}


%%%%%%%%%%%%%%%%%%%%%%%%%%%%%%%%%%
\section{Introduction}

In the Standard Model (SM~\cite{prl_19_1264,np_22_579,sm_salam}), a
scalar field vacuum expectation value breaks the Electroweak symmetry,
gives masses to the $W$ and $Z$
bosons~\cite{prl_13_321,pl_12_132,prl_13_585}, and manifests itself
directly as the so-called Higgs boson. A primary goal of the Large
Hadron Collider (LHC) is to test the SM mechanism of Electroweak
symmetry breaking by searching for Higgs boson production in high
energy proton-proton collisions. Thanks in part to the large gluon
luminosity at LHC energies~\cite{Lai:2010vv,Sherstnev:2007nd}, the
Higgs boson is predominantly produced via gluon fusion 
($gg\rightarrow
H$)~\cite{Anastasiou:2008tj,deFlorian:2009hc,Aglietti:2004nj,Actis:2008ug}
and to a lesser extent via weak boson fusion ($qq\rightarrow
qqH$)~\cite{Bolzoni:2010xr,Ciccolini:2007jr,Ciccolini:2007ec}. Published
limits from direct searches at LEP and the Tevatron exclude $\mH <
114.4$~GeV~\cite{Barate:2003sz} and 162~GeV$< \mH <
166$~GeV~\cite{PhysRevLett.104.061802} at 95\% C.L.

For $m_H \simge 135~\GeV$, the dominant decay mode of the Higgs boson
is $H\rightarrow
WW^{(\ast)}$~\cite{Bredenstein:2007ec,LHCHiggsCrossSectionWorkingGroup:2011ti}.
The most sensitive Higgs boson search channel in the range $135 \simle
m_H \simle 200~\GeV$ is the purely leptonic mode $\HWWstarlvlv$. For
$m_H \simge 200~\GeV$, the $\HWWlvqq$ channel, where one $W$ decays to
a pair of jets ($W \rightarrow jj$), also becomes important, along
with $H\rightarrow ZZ$ channels. The
advantage of $\HWWlvqq$ over $\HWWlvlv$ for a high mass Higgs boson is 
the ability to fully reconstruct the Higgs boson mass.

This document describes a search for a Higgs boson in the
$\HWWlvqq$ channel using the ATLAS detector at the LHC, and based on
1.04 fb$^{-1}$ of $pp$ collisions at $\sqrt{s} = 7$ TeV collected
during 2011~\cite{lvqqPRL}. In this analysis, the distribution of the
$\ell\nu jj$ invariant mass $m(\ell\nu jj)$, reconstructed using the
constraint $m(\ell\nu) = m(W)$ and the requirement that two of the
jets in the event are consistent with a $W\rightarrow jj$ decay, is
used to search for a Higgs boson signal. The results from a previous
search for $H\rightarrow WW\rightarrow\ell\nu jj$ based on 35
pb$^{-1}$ of data collected during the 2010 run of LHC were presented
in Ref.~\cite{Aad:2011qi}.

The present search, based on the measured shape of the $m(\ell\nu jj)$
distribution, is restricted to $m_{H} > 240$ GeV, in order to ensure a
smoothly varying non-resonant background, well clear of the effective
kinematic cutoff $m(\ell\nu jj) \sim 160$ GeV. For $m_H \simge
600~\GeV$, the jets from $W\rightarrow jj$ decay begin to overlap in
the calorimeter, and also the natural width of the Higgs boson becomes
large enough that a more detailed treatment of these issues, beyond
the scope of the present analysis, might be required. The best
sensitivity in this analysis occurs for $m_{H} \sim 400$ GeV. 

%%%%%%%%%%%%%%%%%%%%%%%%%%%%%%%%%%
\section{Detector}

The ATLAS detector~\cite{atlas} is a multipurpose particle
physics apparatus with forward-backward symmetric cylindrical geometry
covering $|\eta|<2.5$ for tracks and $|\eta|<4.9$ for
jets~\footnote{ATLAS uses a right-handed coordinate system with its
  origin at the nominal interaction point (IP) in the center of the
  detector and the $z$-axis coinciding with the axis of the beam
  pipe. The $x$-axis points from the IP to the center of the LHC ring,
  and the $y$-axis points upward. Cylindrical coordinates $(r, \phi)$
  are used in the transverse plane, $\phi$ being the azimuthal angle
  around the beam pipe. The pseudorapidity is defined in terms of the
  polar angle $\theta$ as $\eta = -\ln \tan(\theta/2)$.}.
The inner tracking detector (ID) consists of a silicon pixel detector,
a silicon microstrip detector, and a transition radiation tracker. The
ID is surrounded by a thin superconducting solenoid providing a 2~T
magnetic field, and by a high-granularity liquid-argon (LAr) sampling
electromagnetic (EM) calorimeter. An iron-scintillator tile
calorimeter provides hadronic coverage in the central rapidity
range. The end-cap and forward regions are instrumented with LAr
calorimetry for both electromagnetic and hadronic measurements. The
muon spectrometer surrounds the calorimeters and consists of three
large superconducting toroids, each with eight coils, a system of
precision tracking chambers, and detectors for triggering. 

Detailed Monte Carlo (MC) studies of signal and backgrounds have been
carried out. The interaction with the ATLAS detector is modeled using
GEANT4~\cite{geant4} and the events are reconstructed using the same
software that is used to perform the reconstruction on data.  The
effect of multiple $pp$ interactions occurring at high luminosities
(pile-up) is simulated in the MC samples by superimposing at the
generation stage several simulated minimum-bias events on the
simulated signal and background events. MC samples corresponding to
several different sets of pile-up conditions were generated and
subsequently re-weighted to match the observed pile-up conditions for
the present data set.

%%%%%%%%%%%%%%%%%%%%%%%%%%%%%%%%%%
\section{Dataset and Trigger}

The data used in this analysis were recorded during periods when
stable beams were present, and all ATLAS sub-detectors were operating
under nominal conditions. The events were triggered by requiring the 
presence of an electron candidate with transverse energy
$E_{T}>20$~GeV or a muon candidate with transverse momentum
$p_{T}>18$~GeV.

%%%%%%%%%%%%%%%%%%%%%%%%%%%%%%%%%%
\section{Object Selection}

Electron candidates are selected from clustered energy deposits in the
EM calorimeter with an associated track and are required to satisfy a
tight set of identification cuts~\cite{:2010yt} with an efficiency of
71 $\pm$ 1.6\% for electrons with $E_{T}>20$~GeV. While the energy
measurement is taken from the EM calorimeter, the pseudorapidity
$\eta$ and azimuthal angle $\phi$ are taken from the associated
track. The cluster is required to be in $|\eta|<2.47$, outside
$1.37<|\eta|<1.52$, and in a region where the calorimeter was known to
be operating normally at the time the event was recorded. The track
associated with the electron candidate is required to point back to a
reconstructed primary vertex with a transverse impact parameter
significance $\le 10\sigma$ and an impact parameter along the beam
direction $\le 10$~mm. Electrons are required to be isolated: the sum
of the transverse energies of cells inside a cone $\Delta
R<0.3$\footnote{$\Delta R=\sqrt{\Delta\phi^{2}+\Delta\eta^{2}}$.}
around the cluster centroid (excluding the electron itself) must
satisfy $\Sigma(E_{T}^{calo})<4$~GeV.

Muons are reconstructed by combining tracks in the inner detector and
muon spectrometer\cite{springerlink:10.1007/JHEP12(2010)060}. The
efficiency of this reconstruction is $92\pm 0.6$\% for muons with
$p_{T}>20$~GeV. Muons are required to satisfy basic quality cuts on
the number and type of hits in the inner detector. They must lie in
$|\eta|<2.4$, satisfy the same impact parameter cuts as electrons, and
be isolated, with the sum of track transverse momenta in a cone
$\Delta R<0.2$ around the muon satisfying
$\Sigma(p_{T}^{track})/p_{T}^{\mu}<0.1$.

Jets are reconstructed from calibrated topological clusters using the
anti-$k_{t}$ algorithm~\cite{AntiKt} with radius parameter $R=0.4$.
The reconstructed jets are calibrated using an $E_{T}$ and $\eta$
dependent correction factor based on MC simulation. They are required
to have $E_{T}>25$~GeV and $|\eta|<4.5$. Jets are $b$-tagged if they
contain a reconstructed displaced secondary vertex consistent with a
$b$-decay. The missing transverse energy $\MET$ in the event is
reconstructed starting from topological energy clusters in the
calorimeters calibrated according to the type of the object to which
they belong and accounting also for muons, since muons generally
deposit little of their energy in the calorimeters.

%%%%%%%%%%%%%%%%%%%%%%%%%%%%%%%%%%
\section{Event Selection}

The primary vertex in selected events is required to consist of
$\ge 3$ tracks with $p_{T}>400$~MeV. There must be exactly one
reconstructed lepton candidate (electron or muon) with
$p_{T}>30$~GeV. In order to ensure that this analysis is statistically
independent of the ATLAS $H\rightarrow ZZ\rightarrow\ell\ell\nu\nu$
analysis and thus allow the computation of combined limits in a future
study, events are vetoed if there is any additional lepton with
$p_{T}>20$~GeV, including electrons which only satisfy the looser
identification cuts used in the $H\rightarrow ZZ\rightarrow
\ell\ell\nu\nu$ analysis~\cite{ATLAS-CONF-2011-026}.

Events are required to have $\MET > 30$~GeV consistent with an
unobserved neutrino from $W\rightarrow\ell\nu$ decay. There must be
exactly two jets ($H+0$ jet sample) or exactly three jets ($H+1$ jet
sample) having $E_T > 25$~GeV and $|\eta|<4.5$. The two jets with
invariant mass closest to the mass of the $W$ boson are required to
satisfy $71~\GeV < m_{jj} < 91~\GeV$. These two jets are taken as the
$W$ decay jets and are required to lie in the range $|\eta|<2.8$,
where the jet energy scale (JES) is known to better than $\pm
(4-8)$\% for $E_{T}>25$~GeV.

The mass constraint equation $m(\ell\nu)=m(W)$ is used to reconstruct
$m(\ell\nu jj)$ (taking $p_{T}^{\nu}$ from \MET) and can have real or
complex solutions. In the case of a complex solution, the event is
rejected, while in the case of two real solutions, the solution with
smaller $|p_{z}^{\nu}|$ is taken, based on studies using simulation.

After the event selection, the background is expected to be dominated
by $W$+jets production. Other important backgrounds are $Z$+jets,
multijets (MJ) from QCD processes, top quark and diboson ($WW,WZ$,
and $ZZ$) production. In order to further reject backgrounds from top
quark production, events are rejected if any of the jets passes the
$b$-tagging criteria. The efficiency and mistag rate of the
$b$-tagging have been measured using the methods described in
Ref.~\cite{btaggingCONF} and are well modeled in the MC simulation.

%%%%%%%%%%%%%%%%%%%%%%%%%%%%%%%%%%
\section{Signal and Background Modeling}

Although MC is not used to model the background in the final fit used
to obtain limits, a combination of MC and data-driven models are used
to better understand the background yields at this intermediate
stage. Both the gluon fusion and the weak boson fusion signal
production processes are modeled at NLO using the POWHEG + PYTHIA event
generators~\cite{pythia,powheg,Nason:2009ai}.
Table~\ref{signalCrossSectionTable} shows the cross-sections for the
gluon fusion and weak boson fusion signal processes in the
SM~\cite{LHCHiggsCrossSectionWorkingGroup:2011ti}. 

Backgrounds due to $W/Z$+jets, $t\overline{t}$, and diboson production
are modeled using the ALPGEN~\cite{alpgen}, MC@NLO~\cite{mcatnlo}, and
HERWIG~\cite{herwig} generators, respectively. A small contribution
from $W/Z+\gamma$ events is modeled using
MadEvent~\cite{Alwall:2007st}. The MJ background is modeled using
templates derived from the data selected in an identical way to the 
$H\rightarrow WW\rightarrow\ell\nu jj$ selection except that the
electron identification requirements are loosened (electrons
satisfying the complete set of identification criteria are vetoed) and
the isolation requirement on muons is inverted. Expected contributions
from non-QCD processes to the MJ templates are subtracted using
absolute MC predictions. The normalizations for the MJ and $W/Z$+jets
contributions are determined by a fit at this stage to the $\MET$
distribution with all cuts applied except for the $\MET > 30$~GeV
requirement and with the sum of top and diboson backgrounds fixed to
its expectation in 1.04 fb$^{-1}$ of integrated luminosity.

\begin{table}[ht]
\begin{center}
\caption{Cross-sections for Higgs boson production and the branching
ratio for $H\rightarrow WW\rightarrow\ell\nu jj$ ($\ell=e/\mu$) in the
Standard Model.}
\begin{tabular}{ c | c c | c }
\hline
\hline
$\mH$[GeV]                                      & $\sigma(gg\rightarrow H)$ [pb]        & $\sigma(qq\rightarrow H)$ [pb] & BR($H\rightarrow\ell^{\pm}\nu jj$)\\
\hline
300\rule[-1mm]{0mm}{4.7mm}                      & 2.4$\pm 0.4$                          & 0.30$^{+0.014}_{-0.008}$       & 0.202\\    %0.692\\
400\rule[-1mm]{0mm}{4.7mm}                      & 2.0$^{+0.31}_{-0.34}$                 & 0.162$^{+0.010}_{-0.005}$      & 0.170\\   %0.582\\
500\rule[-1mm]{0mm}{4.7mm}                      & 0.85$\pm 0.15$                        & 0.095$^{+0.0068}_{-0.0032}$    & 0.160\\  %0.546\\
600\rule[-1mm]{0mm}{4.7mm}                      & 0.33$^{+0.063}_{-0.058}$              & 0.058$^{+0.005}_{-0.002}$       & 0.164\\  %0.558\\
\hline
\hline
\end{tabular}
\label{signalCrossSectionTable}
\end{center}
\end{table}

%%%%%%%%%%%%%%%%%%%%%%%%%%%%%%%%%%
\section{Results}

Table~\ref{ggHcutflowTable} shows the observed and expected yields for
signal and background, after the full event selection except for the
requirement that $m(\ell\nu)=m(W)$ has a real solution, for each jet
multiplicity and flavor category and for their sum. 
\begin{table*}
\caption{
Observed and expected event yields in 1.04~fb$^{-1}$ of data, for SM
Higgs boson production and significant backgrounds after all cuts
for the $H+0j$ and $H+1j$ search channels, prior to the requirement
that $m(\ell\nu)=m(W)$ has a real solution. For the $W/Z$+jets and MJ
backgrounds, the uncertainties are taken from a fit to the $\MET$
distribution which normalizes these backgrounds and the total error
in the rightmost column is the sum in quadrature since the fit errors
are statistically independent. For signal, top and diboson production,
the quoted uncertainties are JES ($\pm$17\%), cross-section ($\pm$10\%
for both top and diboson), and luminosity ($\pm$3.7\%) and are added
in quadrature; the total errors in the rightmost column for these 
contributions are the linear sum of the errors for the individual
channels since these sources of systematic uncertainty are correlated
across channels. Statistical errors are small here compared to these
uncertainties.
}
%\begin{ruledtabular}
\begin{tabular}{ c | r r  r r | r }
                           & H($e\nu jj$) + 0j  &  H($\mu\nu jj$) + 0j  &  H($e\nu jj$) + 1j  &  H($\mu\nu jj$) + 1j  &  H + 0j or 1j \\
\hline
$W/Z$+jets                      &   10780 $\pm$ 290       &        13380 $\pm$ 870    &    6510 $\pm$ 250        &    7410 $\pm$ 670         &   38080 $\pm$ 1160 \\
Multi-jet                     &     890 $\pm$  24       &          256 $\pm$  17    &     669 $\pm$  25        &     212 $\pm$ 19          &    2027 $\pm$ 43 \\
Top                           &     170 $\pm$  34       &          164 $\pm$ 33     &     489 $\pm$ 98         &     500 $\pm$ 100         &    1320 $\pm$ 270 \\
Dibosons                      &     397 $\pm$  79       &          414 $\pm$ 83     &     161 $\pm$ 32         &     204 $\pm$ 41          &    1180 $\pm$ 240 \\
\hline
Expected Background           &   12240 $\pm$ 300       &        14210 $\pm$ 870    &    7830 $\pm$ 270        &    8330 $\pm$ 680         &   42600 $\pm$ 1200 \\
\hline
Observed                      &   11988                 &        13906              &     7543                 &    8250                   &   41687     \\
\hline
Expected Signal ($m_H = 400$ GeV)  &     14 $\pm$ 3.6   &        12 $\pm$ 3.1       &       18 $\pm$ 4.7       &       14 $\pm$ 3.6        &      58 $\pm$ 15 \\
\end{tabular}
%\end{ruledtabular}
\label{ggHcutflowTable}
\end{table*}

\begin{figure}[tb]
\begin{center}
\includegraphics[angle=0,width=0.49\textwidth]{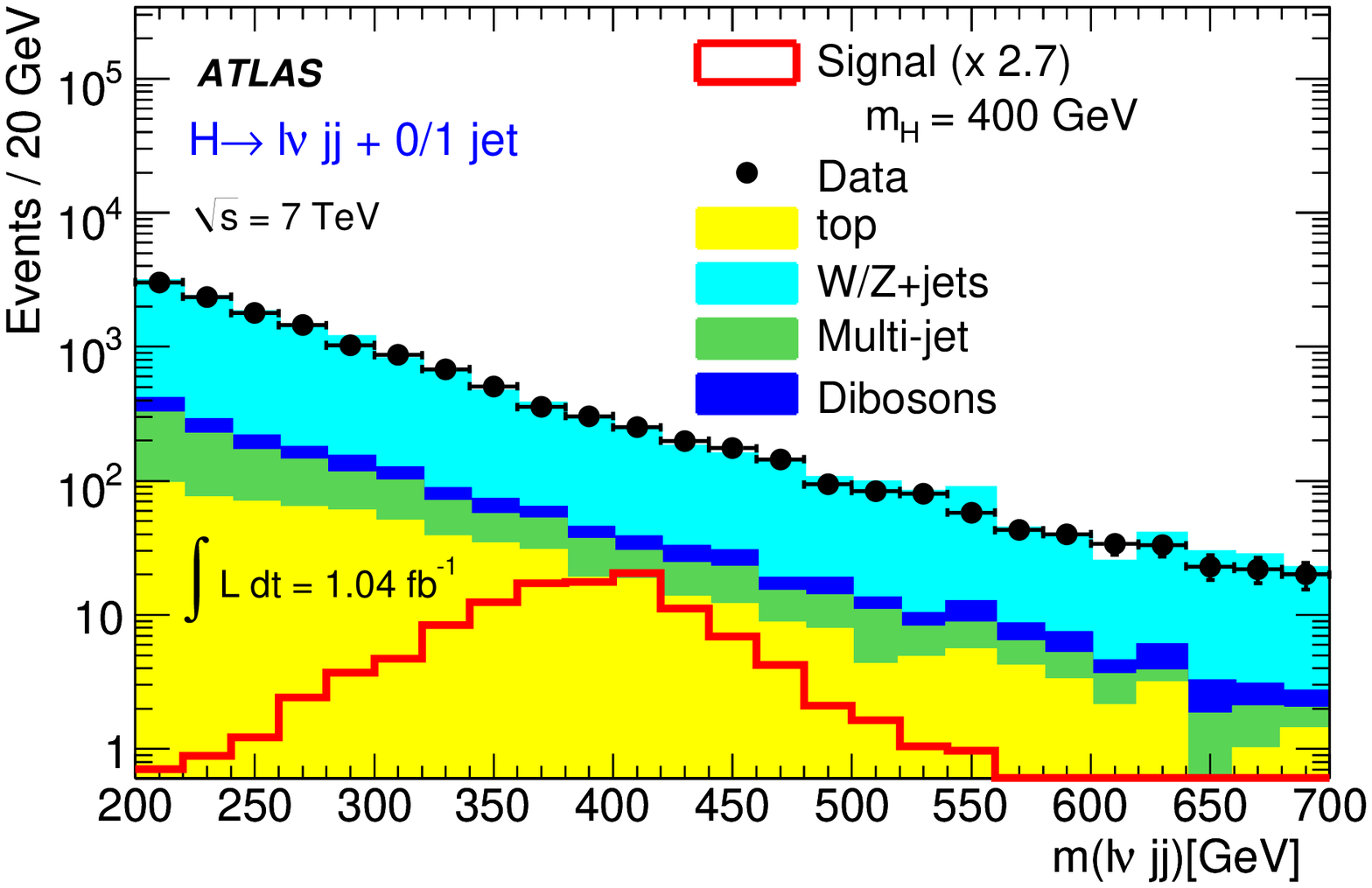}
\includegraphics[angle=0,width=0.49\textwidth]{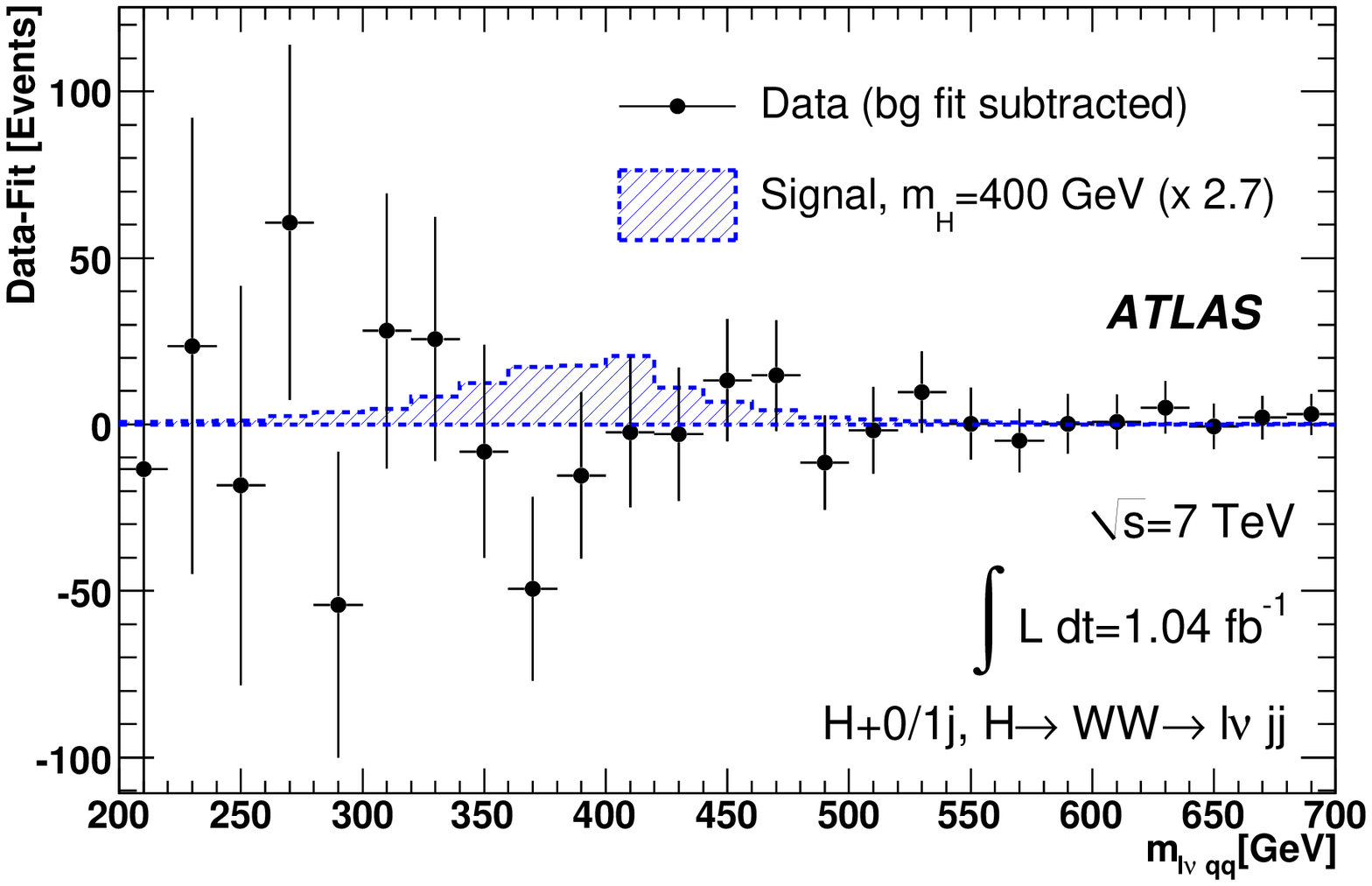}
\caption{
The distribution of the invariant mass $m(\ell\nu jj)$, summed over
lepton flavor and jet multiplicity. Top:  the comparison between data
and MC.  Bottom:  the difference between data and the fitted
non-resonant background. The expected contribution from Higgs boson
decays for $\mH = 400$ GeV in the SM is also shown, multiplied by a
factor of 2.7.} 
\label{mlnuqq01j_plots}
\end{center}
\end{figure}

\begin{figure}[tb]
\begin{center}
\includegraphics[angle=0,width=0.89\textwidth]{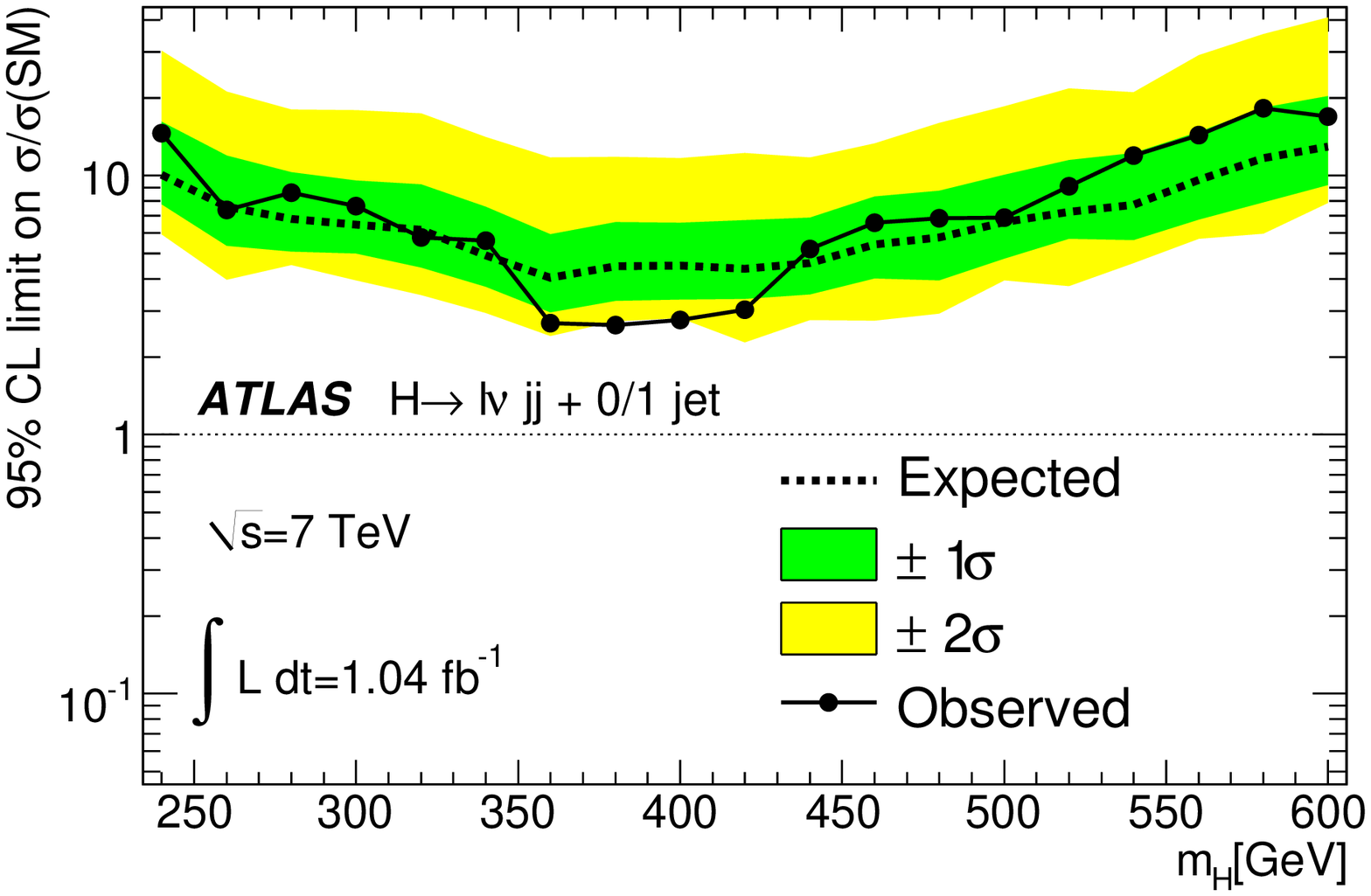}
\caption{
The expected and observed 95\% confidence level upper limits on the
Higgs boson production cross-section ratio with the SM cross-section
in 1.04~fb$^{-1}$ of data. For any hypothesized Higgs boson mass, the
background contribution used in the calculation of this limit is set
by the background distribution obtained from the fit to the $m(\ell\nu
jj)$ distribution.}
\label{limit_plot}
\end{center}
\end{figure}

Figure~\ref{mlnuqq01j_plots} (top) shows the $m(\ell\nu jj)$
distribution, including the requirement that $m(\ell\nu)=m(W)$ has a
real solution, summed over all lepton flavors and jet
multiplicities. The SM signal with $\mH = 400~\GeV$ is also shown,
scaled up by a factor of 2.7.

Limits are set using a maximum likelihood fit to the shape of the
observed $m(\ell\nu jj)$ distribution in the range $200<m(\ell\nu
jj)<2000$~GeV. The non-resonant background in this fit is modeled by
the sum of two exponential functions. The normalization and slope of
each exponential are free parameters in the fit. The
double-exponential form for the total background is well justified by
fits to the $m(\ell\nu jj)$ distributions obtained by selecting events
with $m_{jj}$ just below ($50<m_{jj}<60$~GeV) or just above
($100<m_{jj}<110$~GeV) the $W$ peak, respectively. The $m(\ell\nu jj)$
distribution for the signal at each hypothesized $m_H$ is modeled
using the signal MC samples. 

The fit accounts for the uncertainty in the efficiency of the
electron, muon, and jet reconstruction. The electron and muon
efficiencies are varied within their uncertainties, leading to an
uncertainty in the signal efficiency of $\pm$1.6\% and $\pm$0.6\%, for
electrons and muons respectively. Varying the jet energy scale within
its uncertainties yields a corresponding uncertainty of $\pm$17\%, and
smearing the jet energies within the uncertainty on their resolutions
results in an uncertainty of $\pm$8.6\%. The limits also take into
account a $\pm$3.7\% uncertainty on the luminosity determination and a
$\pm$19.4\% uncertainty on the predicted signal
cross-section~\cite{LHCHiggsCrossSectionWorkingGroup:2011ti}, taken to
be independent of mass.

Effects due to the finite width of the Higgs boson and due to
interference between Higgs boson and SM processes, which were
discussed in
Refs.~\cite{LHCHiggsCrossSectionWorkingGroup:2011ti,Anastasiou:2011pi}, 
have been neglected. If an additional uncertainty of 150$\times
m_{H}^{3}$\% (where $m_{H}$ is in TeV) were to be assigned for this,
it would increase the total systematic error by $<6$\% for $m_{H}\le
500$~GeV, but as much as 15\% for $m_{H}=600$~GeV where the impact on
the limit is about $18$\%.

Figure~\ref{mlnuqq01j_plots} (right) shows the difference between the
$m(\ell\nu jj)$ distribution in data and the fitted background. There
is no indication of any significant excess, so limits are extracted
using the Profile Likelihood~\cite{asymptotics} as a test statistic
and following the $CL_{s}$ procedure described in
Ref.~\cite{cls}. Figure~\ref{limit_plot} shows the 95\% CL upper bound 
on $\sigma\times BR_{H\rightarrow WW}/(\sigma\times BR_{H\rightarrow
  WW})_{\rm SM}$ as a function of $\mH$. The 95\% confidence level in
the data for $\mH = 400~\GeV$ is 3.1 pb, or 2.7 times the SM
prediction. In the Standard model with an additional heavy fourth
generation (SM4)~\cite{Kribs:2007nz,Unel:SM4} the gluon fusion
mechanism for production of a Higgs boson is expected to be
substantially enhanced. Within the SM4 context, a Higgs boson is
excluded at 95\% CL by the present data over the range $m_{H}=310-430$
GeV.

\bigskip % extra skip inserted
\bibliography{dpf2011,Higgs_paper}

\begin{thebibliography}{40}
\expandafter\ifx\csname natexlab\endcsname\relax\def\natexlab#1{#1}\fi
\expandafter\ifx\csname bibnamefont\endcsname\relax
  \def\bibnamefont#1{#1}\fi
\expandafter\ifx\csname bibfnamefont\endcsname\relax
  \def\bibfnamefont#1{#1}\fi
\expandafter\ifx\csname citenamefont\endcsname\relax
  \def\citenamefont#1{#1}\fi
\expandafter\ifx\csname url\endcsname\relax
  \def\url#1{\texttt{#1}}\fi
\expandafter\ifx\csname urlprefix\endcsname\relax\def\urlprefix{URL }\fi
\providecommand{\bibinfo}[2]{#2}
\providecommand{\eprint}[2][]{\url{#2}}

\bibitem[{\citenamefont{Weinberg}(1967)}]{prl_19_1264}
\bibinfo{author}{\bibfnamefont{S.}~\bibnamefont{Weinberg}},
  \bibinfo{journal}{Phys. Rev. Lett.} \textbf{\bibinfo{volume}{19}},
  \bibinfo{pages}{1264} (\bibinfo{year}{1967}).

\bibitem[{\citenamefont{Glashow}(1961)}]{np_22_579}
\bibinfo{author}{\bibfnamefont{S.~L.} \bibnamefont{Glashow}},
  \bibinfo{journal}{Nucl. Phys.} \textbf{\bibinfo{volume}{B22}},
  \bibinfo{pages}{579} (\bibinfo{year}{1961}).

\bibitem[{\citenamefont{Salam}(1968)}]{sm_salam}
\bibinfo{author}{\bibfnamefont{A.}~\bibnamefont{Salam}},
  \emph{\bibinfo{title}{in Elementary Particle Theory, p. 367}}
  (\bibinfo{publisher}{Alm\-qvist and Wiksell}, \bibinfo{address}{Stockholm},
  \bibinfo{year}{1968}).

\bibitem[{\citenamefont{{F.~Englert and R.~Brout}}(1964)}]{prl_13_321}
\bibinfo{author}{\bibnamefont{{F.~Englert and R.~Brout}}},
  \bibinfo{journal}{Phys. Rev. Lett.} \textbf{\bibinfo{volume}{13}},
  \bibinfo{pages}{321} (\bibinfo{year}{1964}).

\bibitem[{\citenamefont{Higgs}(1964)}]{pl_12_132}
\bibinfo{author}{\bibfnamefont{P.~W.} \bibnamefont{Higgs}},
  \bibinfo{journal}{Phys. Lett.} \textbf{\bibinfo{volume}{12}},
  \bibinfo{pages}{132} (\bibinfo{year}{1964}).

\bibitem[{\citenamefont{{G.~S.~Guralnik, C.R.~Hagen and
  T.W.B.~Kibble}}(1964)}]{prl_13_585}
\bibinfo{author}{\bibnamefont{{G.~S.~Guralnik, C.R.~Hagen and T.W.B.~Kibble}}},
  \bibinfo{journal}{Phys. Rev. Lett.} \textbf{\bibinfo{volume}{13}},
  \bibinfo{pages}{585} (\bibinfo{year}{1964}).

\bibitem[{\citenamefont{Lai et~al.}(2010)\citenamefont{Lai, Guzzi, Huston, Li,
  Nadolsky et~al.}}]{Lai:2010vv}
\bibinfo{author}{\bibfnamefont{H.-L.} \bibnamefont{Lai}},
  \bibinfo{author}{\bibfnamefont{M.}~\bibnamefont{Guzzi}},
  \bibinfo{author}{\bibfnamefont{J.}~\bibnamefont{Huston}},
  \bibinfo{author}{\bibfnamefont{Z.}~\bibnamefont{Li}},
  \bibinfo{author}{\bibfnamefont{P.~M.} \bibnamefont{Nadolsky}},
  \bibnamefont{et~al.}, \bibinfo{journal}{Phys.Rev.}
  \textbf{\bibinfo{volume}{D82}}, \bibinfo{pages}{074024}
  (\bibinfo{year}{2010}).

\bibitem[{\citenamefont{Sherstnev and Thorne}(2008)}]{Sherstnev:2007nd}
\bibinfo{author}{\bibfnamefont{A.}~\bibnamefont{Sherstnev}} \bibnamefont{and}
  \bibinfo{author}{\bibfnamefont{R.~S.} \bibnamefont{Thorne}},
  \bibinfo{journal}{Eur. Phys. J.} \textbf{\bibinfo{volume}{C55}},
  \bibinfo{pages}{553} (\bibinfo{year}{2008}).

\bibitem[{\citenamefont{Anastasiou et~al.}(2009)\citenamefont{Anastasiou,
  Boughezal, and Petriello}}]{Anastasiou:2008tj}
\bibinfo{author}{\bibfnamefont{C.}~\bibnamefont{Anastasiou}},
  \bibinfo{author}{\bibfnamefont{R.}~\bibnamefont{Boughezal}},
  \bibnamefont{and}
  \bibinfo{author}{\bibfnamefont{F.}~\bibnamefont{Petriello}},
  \bibinfo{journal}{JHEP} \textbf{\bibinfo{volume}{04}}, \bibinfo{pages}{003}
  (\bibinfo{year}{2009}).

\bibitem[{\citenamefont{de~Florian and Grazzini}(2009)}]{deFlorian:2009hc}
\bibinfo{author}{\bibfnamefont{D.}~\bibnamefont{de~Florian}} \bibnamefont{and}
  \bibinfo{author}{\bibfnamefont{M.}~\bibnamefont{Grazzini}},
  \bibinfo{journal}{Phys. Lett.} \textbf{\bibinfo{volume}{B674}},
  \bibinfo{pages}{291} (\bibinfo{year}{2009}).

\bibitem[{\citenamefont{Aglietti et~al.}(2004)\citenamefont{Aglietti, Bonciani,
  Degrassi, and Vicini}}]{Aglietti:2004nj}
\bibinfo{author}{\bibfnamefont{U.}~\bibnamefont{Aglietti}},
  \bibinfo{author}{\bibfnamefont{R.}~\bibnamefont{Bonciani}},
  \bibinfo{author}{\bibfnamefont{G.}~\bibnamefont{Degrassi}}, \bibnamefont{and}
  \bibinfo{author}{\bibfnamefont{A.}~\bibnamefont{Vicini}},
  \bibinfo{journal}{Phys. Lett.} \textbf{\bibinfo{volume}{B595}},
  \bibinfo{pages}{432} (\bibinfo{year}{2004}).

\bibitem[{\citenamefont{Actis et~al.}(2008)\citenamefont{Actis, Passarino,
  Sturm, and Uccirati}}]{Actis:2008ug}
\bibinfo{author}{\bibfnamefont{S.}~\bibnamefont{Actis}},
  \bibinfo{author}{\bibfnamefont{G.}~\bibnamefont{Passarino}},
  \bibinfo{author}{\bibfnamefont{C.}~\bibnamefont{Sturm}}, \bibnamefont{and}
  \bibinfo{author}{\bibfnamefont{S.}~\bibnamefont{Uccirati}},
  \bibinfo{journal}{Phys. Lett.} \textbf{\bibinfo{volume}{B670}},
  \bibinfo{pages}{12} (\bibinfo{year}{2008}).

\bibitem[{\citenamefont{Bolzoni et~al.}(2010)\citenamefont{Bolzoni, Maltoni,
  Moch, and Zaro}}]{Bolzoni:2010xr}
\bibinfo{author}{\bibfnamefont{P.}~\bibnamefont{Bolzoni}},
  \bibinfo{author}{\bibfnamefont{F.}~\bibnamefont{Maltoni}},
  \bibinfo{author}{\bibfnamefont{S.-O.} \bibnamefont{Moch}}, \bibnamefont{and}
  \bibinfo{author}{\bibfnamefont{M.}~\bibnamefont{Zaro}},
  \bibinfo{journal}{Phys. Rev. Lett.} \textbf{\bibinfo{volume}{105}},
  \bibinfo{pages}{011801} (\bibinfo{year}{2010}), \eprint{1003.4451}.

\bibitem[{\citenamefont{Ciccolini et~al.}(2007)\citenamefont{Ciccolini, Denner,
  and Dittmaier}}]{Ciccolini:2007jr}
\bibinfo{author}{\bibfnamefont{M.}~\bibnamefont{Ciccolini}},
  \bibinfo{author}{\bibfnamefont{A.}~\bibnamefont{Denner}}, \bibnamefont{and}
  \bibinfo{author}{\bibfnamefont{S.}~\bibnamefont{Dittmaier}},
  \bibinfo{journal}{Phys. Rev. Lett.} \textbf{\bibinfo{volume}{99}},
  \bibinfo{pages}{161803} (\bibinfo{year}{2007}), \eprint{0707.0381}.

\bibitem[{\citenamefont{Ciccolini et~al.}(2008)\citenamefont{Ciccolini, Denner,
  and Dittmaier}}]{Ciccolini:2007ec}
\bibinfo{author}{\bibfnamefont{M.}~\bibnamefont{Ciccolini}},
  \bibinfo{author}{\bibfnamefont{A.}~\bibnamefont{Denner}}, \bibnamefont{and}
  \bibinfo{author}{\bibfnamefont{S.}~\bibnamefont{Dittmaier}},
  \bibinfo{journal}{Phys. Rev.} \textbf{\bibinfo{volume}{D77}},
  \bibinfo{pages}{013002} (\bibinfo{year}{2008}), \eprint{0710.4749}.

\bibitem[{\citenamefont{{The LEP Collaborations}}(2003)}]{Barate:2003sz}
\bibinfo{author}{\bibnamefont{{The LEP Collaborations}}},
  \bibinfo{journal}{Phys. Lett.} \textbf{\bibinfo{volume}{B565}},
  \bibinfo{pages}{61} (\bibinfo{year}{2003}).

\bibitem[{\citenamefont{Aaltonen et~al.}(2010)}]{PhysRevLett.104.061802}
\bibinfo{author}{\bibfnamefont{T.}~\bibnamefont{Aaltonen}} \bibnamefont{et~al.}
  (\bibinfo{collaboration}{CDF Collaboration}), \bibinfo{journal}{Phys. Rev.
  Lett.} \textbf{\bibinfo{volume}{104}}, \bibinfo{pages}{061802}
  (\bibinfo{year}{2010}).

\bibitem[{\citenamefont{Bredenstein et~al.}(2007)\citenamefont{Bredenstein,
  Denner, Dittmaier, and Weber}}]{Bredenstein:2007ec}
\bibinfo{author}{\bibfnamefont{A.}~\bibnamefont{Bredenstein}},
  \bibinfo{author}{\bibfnamefont{A.}~\bibnamefont{Denner}},
  \bibinfo{author}{\bibfnamefont{S.}~\bibnamefont{Dittmaier}},
  \bibnamefont{and} \bibinfo{author}{\bibfnamefont{M.~M.} \bibnamefont{Weber}},
  \bibinfo{journal}{JHEP} \textbf{\bibinfo{volume}{0702}}, \bibinfo{pages}{080}
  (\bibinfo{year}{2007}).

\bibitem[{\citenamefont{{LHC Higgs Cross Section Working Group} et~al.}(CERN,
  Geneva, 2011)\citenamefont{{LHC Higgs Cross Section Working Group},
  Dittmaier, Mariotti, Passarino, and
  Tanaka~(Eds.)}}]{LHCHiggsCrossSectionWorkingGroup:2011ti}
\bibinfo{author}{\bibnamefont{{LHC Higgs Cross Section Working Group}}},
  \bibinfo{author}{\bibfnamefont{S.}~\bibnamefont{Dittmaier}},
  \bibinfo{author}{\bibfnamefont{C.}~\bibnamefont{Mariotti}},
  \bibinfo{author}{\bibfnamefont{G.}~\bibnamefont{Passarino}},
  \bibnamefont{and}
  \bibinfo{author}{\bibfnamefont{R.}~\bibnamefont{Tanaka~(Eds.)}},
  \bibinfo{journal}{CERN-2011-002}  (\bibinfo{year}{CERN, Geneva, 2011}),
  \eprint{arXiv:1101.0593}.

\bibitem[{\citenamefont{{The ATLAS
  Collaboration}}(2011{\natexlab{a}})}]{lvqqPRL}
\bibinfo{author}{\bibnamefont{{The ATLAS Collaboration}}}
  (\bibinfo{year}{2011}{\natexlab{a}}), \eprint{arXiV:1109.3615, submitted to
  {\it Phys. Rev. Lett.}}

\bibitem[{\citenamefont{{The ATLAS
  Collaboration}}(2011{\natexlab{b}})}]{Aad:2011qi}
\bibinfo{author}{\bibnamefont{{The ATLAS Collaboration}}}
  (\bibinfo{year}{2011}{\natexlab{b}}), \eprint{arXiv:1106.2748, submitted to
  EPJC}.

\bibitem[{\citenamefont{{The ATLAS Collaboration}}(2008)}]{atlas}
\bibinfo{author}{\bibnamefont{{The ATLAS Collaboration}}},
  \bibinfo{journal}{JINST} \textbf{\bibinfo{volume}{3}},
  \bibinfo{pages}{S08003} (\bibinfo{year}{2008}).

\bibitem[{\citenamefont{{S.~Agostinelli et al.}}(2003)}]{geant4}
\bibinfo{author}{\bibnamefont{{S.~Agostinelli et al.}}}, \bibinfo{journal}{NIM
  A} \textbf{\bibinfo{volume}{506}}, \bibinfo{pages}{250}
  (\bibinfo{year}{2003}).

\bibitem[{\citenamefont{{The ATLAS
  Collaboration}}(2010{\natexlab{a}})}]{:2010yt}
\bibinfo{author}{\bibnamefont{{The ATLAS Collaboration}}},
  \bibinfo{journal}{JHEP} \textbf{\bibinfo{volume}{12}}, \bibinfo{pages}{060}
  (\bibinfo{year}{2010}{\natexlab{a}}).

\bibitem[{\citenamefont{{The ATLAS
  Collaboration}}(2010{\natexlab{b}})}]{springerlink:10.1007/JHEP12(2010)060}
\bibinfo{author}{\bibnamefont{{The ATLAS Collaboration}}},
  \bibinfo{journal}{JHEP} \textbf{\bibinfo{volume}{2010}}, \bibinfo{pages}{1}
  (\bibinfo{year}{2010}{\natexlab{b}}).

\bibitem[{\citenamefont{Cacciari et~al.}(2008)\citenamefont{Cacciari, Salam,
  and Soyez}}]{AntiKt}
\bibinfo{author}{\bibfnamefont{M.}~\bibnamefont{Cacciari}},
  \bibinfo{author}{\bibfnamefont{G.~P.} \bibnamefont{Salam}}, \bibnamefont{and}
  \bibinfo{author}{\bibfnamefont{G.}~\bibnamefont{Soyez}},
  \bibinfo{journal}{JHEP} \textbf{\bibinfo{volume}{04}}, \bibinfo{pages}{063}
  (\bibinfo{year}{2008}).

\bibitem[{\citenamefont{{The ATLAS
  Collaboration}}(2011{\natexlab{c}})}]{ATLAS-CONF-2011-026}
\bibinfo{author}{\bibnamefont{{The ATLAS Collaboration}}},
  \bibinfo{howpublished}{{ATLAS Conference Note ATLAS-CONF-2011-026, available
  online at http://cdsweb.cern.ch/record/1336759}}
  (\bibinfo{year}{2011}{\natexlab{c}}).

\bibitem[{\citenamefont{{The ATLAS
  Collaboration}}(2011{\natexlab{d}})}]{btaggingCONF}
\bibinfo{author}{\bibnamefont{{The ATLAS Collaboration}}},
  \bibinfo{howpublished}{{ATLAS Conference Note ATLAS-CONF-2011-089, available
  online at http://cdsweb.cern.ch/record/1356198}}
  (\bibinfo{year}{2011}{\natexlab{d}}).

\bibitem[{\citenamefont{{T. Sj\"ostrand et al.}}(2006)}]{pythia}
\bibinfo{author}{\bibnamefont{{T. Sj\"ostrand et al.}}},
  \bibinfo{journal}{JHEP} \textbf{\bibinfo{volume}{05}}, \bibinfo{pages}{026}
  (\bibinfo{year}{2006}).

\bibitem[{\citenamefont{{S. Alioli {\it et. al}}}(2009)}]{powheg}
\bibinfo{author}{\bibnamefont{{S. Alioli {\it et. al}}}},
  \bibinfo{journal}{JHEP} \textbf{\bibinfo{volume}{04}}, \bibinfo{pages}{002}
  (\bibinfo{year}{2009}).

\bibitem[{\citenamefont{Nason and Oleari}(2010)}]{Nason:2009ai}
\bibinfo{author}{\bibfnamefont{P.}~\bibnamefont{Nason}} \bibnamefont{and}
  \bibinfo{author}{\bibfnamefont{C.}~\bibnamefont{Oleari}},
  \bibinfo{journal}{JHEP} \textbf{\bibinfo{volume}{02}}, \bibinfo{pages}{037}
  (\bibinfo{year}{2010}), \eprint{0911.5299}.

\bibitem[{\citenamefont{Mangano et~al.}(2003)}]{alpgen}
\bibinfo{author}{\bibfnamefont{M.~L.} \bibnamefont{Mangano}}
  \bibnamefont{et~al.}, \bibinfo{journal}{JHEP} \textbf{\bibinfo{volume}{07}},
  \bibinfo{pages}{001} (\bibinfo{year}{2003}).

\bibitem[{\citenamefont{Frixione and Webber}(2003)}]{mcatnlo}
\bibinfo{author}{\bibfnamefont{S.}~\bibnamefont{Frixione}} \bibnamefont{and}
  \bibinfo{author}{\bibfnamefont{B.}~\bibnamefont{Webber}},
  \bibinfo{journal}{JHEP} \textbf{\bibinfo{volume}{0308}}, \bibinfo{pages}{007}
  (\bibinfo{year}{2003}).

\bibitem[{\citenamefont{Corcella et~al.}(2001)}]{herwig}
\bibinfo{author}{\bibfnamefont{G.}~\bibnamefont{Corcella}}
  \bibnamefont{et~al.}, \bibinfo{journal}{JHEP} \textbf{\bibinfo{volume}{01}},
  \bibinfo{pages}{010} (\bibinfo{year}{2001}).

\bibitem[{\citenamefont{Alwall et~al.}(2007)}]{Alwall:2007st}
\bibinfo{author}{\bibfnamefont{J.}~\bibnamefont{Alwall}} \bibnamefont{et~al.},
  \bibinfo{journal}{JHEP} \textbf{\bibinfo{volume}{09}}, \bibinfo{pages}{028}
  (\bibinfo{year}{2007}).

\bibitem[{\citenamefont{Anastasiou et~al.}(2011)\citenamefont{Anastasiou,
  Buehler, Herzog, and Lazopoulos}}]{Anastasiou:2011pi}
\bibinfo{author}{\bibfnamefont{C.}~\bibnamefont{Anastasiou}},
  \bibinfo{author}{\bibfnamefont{S.}~\bibnamefont{Buehler}},
  \bibinfo{author}{\bibfnamefont{F.}~\bibnamefont{Herzog}}, \bibnamefont{and}
  \bibinfo{author}{\bibfnamefont{A.}~\bibnamefont{Lazopoulos}}
  (\bibinfo{year}{2011}), \eprint{arXiv:1107.0683}.

\bibitem[{\citenamefont{Cowan et~al.}(2011)}]{asymptotics}
\bibinfo{author}{\bibfnamefont{G.}~\bibnamefont{Cowan}} \bibnamefont{et~al.},
  \bibinfo{journal}{Eur. Phys. J.} \textbf{\bibinfo{volume}{C71}},
  \bibinfo{pages}{1554} (\bibinfo{year}{2011}).

\bibitem[{\citenamefont{Read}(2000)}]{cls}
\bibinfo{author}{\bibfnamefont{A.~L.} \bibnamefont{Read}},
  \emph{\bibinfo{title}{{Modified frequentist analysis of search results (the
  $CL_s$ method)}}}, \bibinfo{howpublished}{{CERN-OPEN-2000-205, available
  online at http://cdsweb.cern.ch/record/451614}} (\bibinfo{year}{2000}).

\bibitem[{\citenamefont{Kribs et~al.}(2007)\citenamefont{Kribs, Plehn,
  Spannowsky, and Tait}}]{Kribs:2007nz}
\bibinfo{author}{\bibfnamefont{G.~D.} \bibnamefont{Kribs}},
  \bibinfo{author}{\bibfnamefont{T.}~\bibnamefont{Plehn}},
  \bibinfo{author}{\bibfnamefont{M.}~\bibnamefont{Spannowsky}},
  \bibnamefont{and} \bibinfo{author}{\bibfnamefont{T.~M.} \bibnamefont{Tait}},
  \bibinfo{journal}{Phys.Rev.} \textbf{\bibinfo{volume}{D76}},
  \bibinfo{pages}{075016} (\bibinfo{year}{2007}), \eprint{0706.3718}.

\bibitem[{\citenamefont{{Becerici Schmidt, N.}
  et~al.}(2010)\citenamefont{{Becerici Schmidt, N.}, {\c{C}etin, S. A.},
  {Istin, S.}, and {Sultansoy, S.}}}]{Unel:SM4}
\bibinfo{author}{\bibnamefont{{Becerici Schmidt, N.}}},
  \bibinfo{author}{\bibnamefont{{\c{C}etin, S. A.}}},
  \bibinfo{author}{\bibnamefont{{Istin, S.}}}, \bibnamefont{and}
  \bibinfo{author}{\bibnamefont{{Sultansoy, S.}}}, \bibinfo{journal}{Eur. Phys.
  J. C} \textbf{\bibinfo{volume}{66}}, \bibinfo{pages}{119}
  (\bibinfo{year}{2010}).

\end{thebibliography}

\end{document}